# Hitchhiker's Guide to Cancer-Associated Lymphoid Aggregates in Histology Images: Manual and Deep Learning-Based Quantification Approaches

Karina Silina[1]*, Francesco Ciompi[2]*

[1] Institute of Pharmaceutical Sciences, Department of Chemistry and Applied Biosciences, Swiss Federal Institute of Technology (ETHZ), Zurich, Switzerland
[2] Pathology Department, Radboud University Medical Center, Nijmegen, Netherlands
* Correspondence to: karina.silina@pharma.ethz.ch, Francesco.Ciompi@radboudumc.nl

## Abstract
Quantification of lymphoid aggregates including tertiary lymphoid structures with germinal centers in histology images of cancer is a promising approach for developing prognostic and predictive tissue biomarkers. In this article, we provide recommendations for identifying lymphoid aggregates in tissue sections from routine pathology workflows such as hematoxylin and eosin staining. To overcome the intrinsic variability associated with manual image analysis (such as subjective decision making, attention span), we recently developed a deep learning-based algorithm called HookNet-TLS to detect lymphoid aggregates and germinal centers in various tissues. Here, we additionally provide a guideline for using manually annotated images for training and implementing HookNet-TLS for automated and objective quantification of lymphoid aggregates in various cancer types.

## 1. Introduction

The tumor microenvironment comprises a variety of infiltrating immune cells **(1)** following two main distribution patterns – scattered throughout the tissue parenchyma, or accumulated in dense focal niches also referred to as immune cell aggregates or clusters. The number and composition of such clusters can differ widely across different tumor types and patients **(2)**. Clusters containing a major B cell component are called tertiary lymphoid structures (TLS), which develop via the process of lymphoid neogenesis involving the interactions between dendritic cells, T cells, B cells and local fibroblasts **(2, 3)**. Similarly to secondary lymphoid organs, priming of antigen specific T-cells and ectopic germinal center reactions take place in TLS **(4, 5)**. It is suggested that these features underlie their often-observed positive prognostic and predictive associations **(3, 6–8)**. Besides the canonical TLS, immune cell aggregates with few B cells and a dominant fraction of T cells and dendritic cells have also been reported **(9–12)**. Whether such aggregates represent stages of canonical lymphoid neogenesis or develop independently of it, is currently not well understood. Importantly, TLS and T cell aggregates are the predominant sites where important predictors of effective anti-tumor immunity such as PD-1$^{hi}$CXCL13$^+$ **(13)** T cells and the so-called progenitor- or stem-like T cells **(10–12, 14, 15)** (identified as PD-1$^+$TCF1$^+$) accumulate, which may represent an additional functional mechanism contributing to augmented anti-tumor immunity *(16)*. Further research is warranted to understand what factors define the heterogeneity of immune cell niche development and function observed across various tumor types *(2)*. This calls for the establishment of unified histological definitions and harmonization of quantification methods of lymphoid aggregates across studies.

Measuring transcriptional profiles as surrogates for lymphoid neogenesis have been proposed in cancer *(3)*. Our analysis of bladder cancer cohort from the Cancer Genome Atlas, however, indicates that the correlation between the transcripts measured in tumor core biopsies and histologically defined numbers of lymphoid aggregates that predominantly form at the tumor invasive front is low *(10)*. Thus, we consider histological analysis as the most reliable approach to quantify dense immune cell niches as spatially identifiable microanatomical objects. Pathology archives of clinical centers are a rich source of histological material, which is routinely analyzed by hematoxylin and eosin (H&E) staining and immunostaining. In recent years, the rise of digital pathology has enabled the transition from glass to high resolution whole-slide images (WSI) that can be used for both clinical and research purposes. The excellent morphological detail, ease of access and the low cost are crucial aspects for the choice of H&E slides as materials for studying lymphoid aggregates in multiple studies *(2)*. In this article, we focus on the analysis of digital pathology images and discuss the strengths and limitations of H&E-based analysis. We provide guidelines for defining lymphoid aggregates in H&E images by manual assessment as well as for training and validating a recently presented deep-learning algorithm, HookNet-TLS *(9)*, for use in various organs.

We refer to dense lymphoid structures as aggregates or clusters through-out the Methods section as it is not possible to discern the morphology of B cells and T cells by human eye in H&E staining *(9)*. This prevents the finer classification of lymphoid aggregates into T cell- or B cell-dominant structures described above. However, mature TLS with fully formed germinal centers can be readily identified in H&E images based on the distinct morphological features of naïve lymphocytes and proliferating germinal center blasts *(7, 9)*.

Another aspect to consider is the sensitivity of aggregate detection: this does not differ between different immunostaining approaches such as immunohistochemistry and multiplex immunofluorescence, while immunostaining shows higher sensitivity than H&E staining (**Figure 1A-B**), especially for aggregates of small size *(9)*. Nevertheless, aggregate density or counts determined in H&E slides have provided reproducible and relevant prognostic and predictive correlations across multiple tumor types *(7, 9, 17)*.

We recommend using images of whole sections over tissue microarrays as well as surgical resections over biopsies whenever possible as the size of the tissue area significantly affects the sensitivity of detection of such aggregates (**Figure 1C-D**). Furthermore, TLS develop mainly in peritumoral areas of various solid tumors including bladder cancer *(10, 18)*, breast cancer *(19–21)*, colorectal cancer *(22–25)*, lung cancer *(17, 26)* and others, which are often omitted in tissue microarrays and are poorly represented in biopsies. It is also important to keep in mind the organ-specific patterns of lymphoid neogenesis to evaluate whether the image contains the appropriate areas for analysis. For example, if working with brain samples the presence of meningeal areas is crucial as it represents the most common region of TLS development in the brain *(12, 27)*.

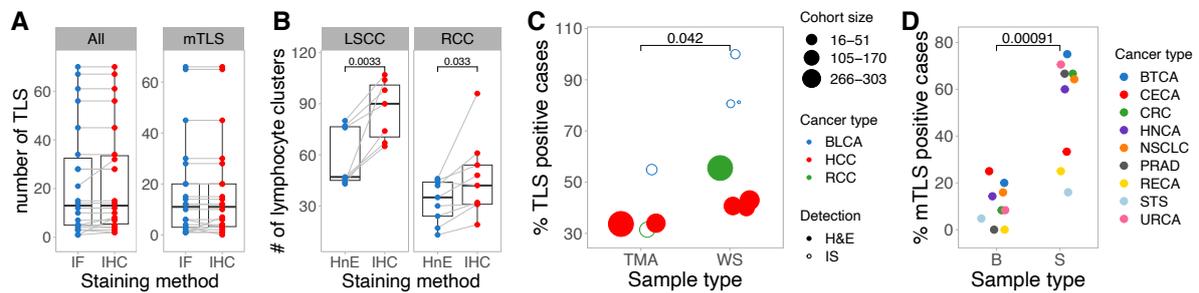

Figure 1. Sensitivity of TLS detection in different staining methods and sample types.
**A** Analysis of data from Supplementary Table 5 of Vanhersheke et al. *(7)*. The number of TLS was detected in serially stained sections of immunohistochemistry (IHC) or immunofluorescence (IF) *(7)*. Surgical resection samples of various tumor types (n=22) were selected for the analysis. The number of TLS (All, all TLS; mTLS, mature TLS) was compared between serially stained slides using the two staining methods by paired t test. **B** Serial sections stained by H&E and IHC for CD20 of the USZ cohort from van Rijthoven et al. *(9)* were processed according the here described Methods section 3.1 steps 1-4. The obtained lymphoid aggregate counts were compared between serially stained slides using the two staining methods by paired t test. LSCC, lung squamous cell carcinoma; RCC, renal clear cell carcinoma. **C** Comparison of TLS detection sensitivity in tissue micro arrays (TMA) versus whole sections (WS). Published studies where histological assessment of TLS was performed in the most common tumor types were explored (n=62). Among identified unique cohorts of untreated patients with reported numbers of TLS-positive cases (n=101) 11 cohorts were selected representing tumor types where matching tissue regions were analyzed in both sample types –TMA and WS (Table 1). Filled circles indicate use of H&E and empty circles – immunostaining (IS). Frequency of TLS-positive cases detected in TMA versus WS samples from these cohorts was compared by Mann-Whitney U test. BLCA, bladder cancer; HCC, hepatocellular carcinoma; RCC, renal clear cell carcinoma. **D** Comparison of TLS detection sensitivity in surgical resections (S) versus biopsies (B). Analysis of histological data from the discovery cohort (n=328) from *(7)*. Presence of mature TLS (mTLS) was determined in various tumor types by multiplex immunofluorescence *(7)*. Frequency of TLS-positive cases was compared between B and S sample types by Mann-Whitney U test in tumor types with more than three cases per sample type. BTCA, biliary tract cancer; CECA, cervical carcinoma; CRC, colorectal cancer; HNCA, head and neck squamous cell carcinoma; NSCLC, non-small cell lung cancer; PRAD, prostate adenocarcinoma; RECA, renal cancer; STS, soft-tissue sarcoma; URCA, urothelial carcinoma.

Table 1. Selected unique cohorts of untreated patients with matched analyzed tissue region in each tumor type but distinct sample type.
BLCA, bladder cancer; HCC, hepatocellular carcinoma; RCC, renal clear cell carcinoma; WS, whole section; TMA, tissue microarray; IM, invasive margin; TC, tumor center.

| Cancer type | # of cases analyzed | # of TLS⁺ cases | % TLS⁺ cases | Sample type | Staining method | Assessed tissue region | Reference |
|---|---|---|---|---|---|---|---|
| BLCA | 16 | 13 | 81 | WS | IS | whole tissue area | *(33)* |
| BLCA | 31 | 25 | 81 | WS | IS | whole tissue area | *(18)* |
| BLCA | 51 | 28 | 55 | TMA | IS | TMA cores of IM and TC | *(34)* |
| BLCA | 40 | 40 | 100 | WS | IS | whole tissue area | *(10)* |
| HCC | 120 | 48 | 40 | WS | H&E | intratumoral area | *(35)* |
| HCC | 159 | 54 | 34 | TMA | H&E | intratumoral area | *(36)* |

| HCC | 145 | 59 | 41 | WS | H&E | intratumoral area | *(35)* |
| HCC | 170 | 73 | 43 | WS | H&E | intratumoral area | *(37)* |
| HCC | 303 | 102 | 34 | TMA | H&E | intratumoral area | *(36)* |
| RCC | 105 | 33 | 31 | TMA | IS | TMA cores of IM and TC | *(34)* |
| RCC | 299 | 166 | 56 | WS | H&E | whole tissue area | *(9)* |

Mucosal organs, such as gut, nasopharynx, tongue and others possess homeostatic mucosa-associated lymphoid tissue (MALT) *(28)*. The assessment of isolated lymphoid follicles (homeostatic MALTs of the gut) was reported to affect the status of *de novo* follicles and provide prognostic information in colorectal cancer *(29)*. Nevertheless, to faithfully assess tumor-induced lymphoid neogenesis in mucosal tissues, it is necessary to distinguish the *de novo* formed lymphoid clusters from the pre-existing homeostatic MALTs. Similarly, lymphoid neogenesis cannot be faithfully assessed by H&E in secondary lymphoid organs (such as in cases of nodal metastases) as their high histological similarity prevents distinguishing *de novo* formed follicles from remaining lymph node parenchyma. Some tumors like non-small cell lung cancer (NSCLC) may also reach and invade draining lymph nodes (LN) *per continuitatem*, which also need to be distinguished from *de novo* aggregates in histology images.

Manual (by eye) assessment of lymphoid aggregates in H&E images involves subjective decision making leading to some level of discrepancy among evaluators (quantitatively demonstrated by extensive histological analysis of TLS in *(7)*). Furthermore, discrepancies in TLS definitions cause the largest variation in aggregate quantification among published studies *(2)*. To address this issue, we developed a deep learning algorithm called HookNet-TLS *(9)* for objective and automated quantification of lymphoid aggregates of various tissues. We found that defining a minimal aggregate size was difficult to implement in manual assessment of H&E images, especially in highly inflamed cases. We thus base our decision mainly on cellular density features: (1) observable difference in cell density delineating the border between a lymphoid aggregate and surrounding parenchyma, (2) the space between lymphocytes in an aggregate is either not visible (in areas with very high density of cells) or clearly differs from surrounding parenchyma (when cell density in a cluster is lower) (**Figure 2A**). These density features allow aggregate identification already at lower magnifications (**Figure 2B**). The shapes of such aggregates are mainly regular (circular/oval); however, also irregular shapes are counted if meet the density criteria (**Figure 2C,** arrowhead). To define germinal centers, we look for a region of morphologically larger cells and lower cell density within the annotated lymphoid aggregates (**Figure 2D**). Finally, aggregates containing a high proportion of plasma cells or non-lymphoid cells are not considered as part of lymphoid neogenesis.

The HookNet-TLS algorithm learns from manually-annotated regions of whole slide images using the so-called multi-resolution architecture, which entails assessment of image data collected at both low and high resolutions (i.e., slide magnification), mimicking the zooming-in and zooming-out procedure of a pathologist's workflow *(9)* (exemplified in **Figure 2**). This way the algorithm captures information of the surrounding tissue context as well as the high-resolution details, which allow the learning and recognition of the morphological features of aggregates and germinal centers described above. HookNet-TLS can be trained using manually annotated images to detect lymphoid aggregates and germinal centers by automatically recognizing and segmenting (i.e., delineating their borders) their morphological characteristics within a variety of different organ parenchymal regions.

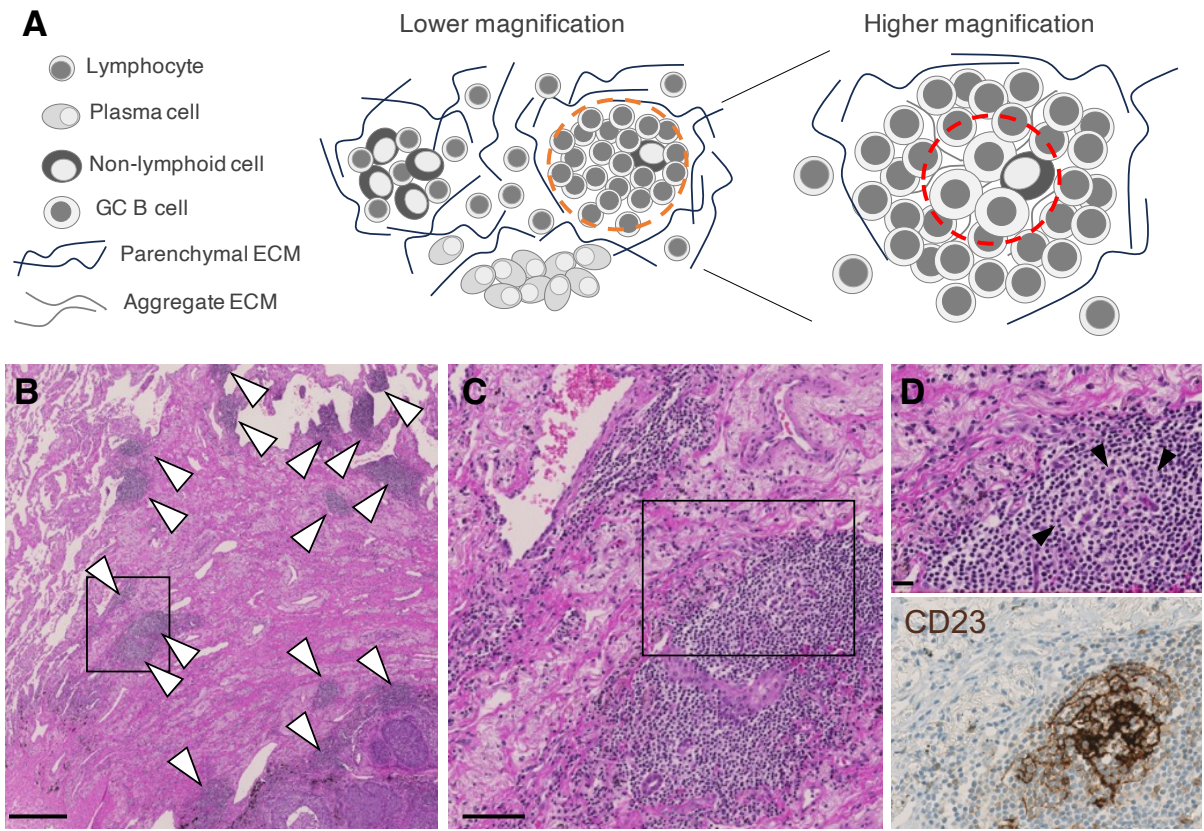

**Figure 2. Example of lymphoid aggregate identification in lung squamous cell carcinoma.**
**A** A schematic representation of histological features at low (left) and high (right) magnification governing lymphoid aggregate (orange circle) and germinal center (red circle) definitions in H&E images. Only clusters consisting of mainly lymphocytes are considered here to be part of the lymphoid neogenesis process. **B** Example of a low magnification histology image of LSCC tissue. Tumor cells occupy the lower right corner of the image; multiple dense lymphocytic clusters can be recognized in the invasive front and immediate adjacent normal lung (arrowheads). Scale bar = 500 μm. **C** Magnification of the area boxed in B. Clear difference in cellular density can be recognized between lymphoid aggregates and surrounding parenchyma. Scale bar = 100 μm. **D** Magnification of the boxed area in C (top). The stroma within the aggregate (the space between lymphocytes) shows differences in extracellular matrix staining, which appears lighter in comparison to the surrounding parenchyma. The reticular network of aggregate-shaping stromal cells can be seen (arrowheads). A circular region with lower cell density can be seen within the aggregate corresponding to a germinal center area, corroborated by immunostaining of CD23 (bottom). Scale bar = 20 μm

## 2. Materials
- Whole slide images of H&E-stained tissues scanned at the resolution of 40X (**Notes 1 and 2**).
- QuPath software *(30)* (open access, Mac/Windows compatible).
- ImageJ software *(31)* (open access, Mac/Windows compatible).

- R (https://www.R-project.org/) and RStudio (http://www.rstudio.com/) (open access, Mac/Windows compatible).
- Script for quantitative data export from QuPath annotations "MeasureAnnotationAreas.ijm" (**Box 1**).
- Script for summarizing the data of the exported manual annotations "Summarize.Rmd", (**Box 2**).
- Script for converting QuPath annotations into .json files for training and testing of HookNet-TLS "json_export.groovy" (**Box 3**) (**Note 3**).
- HookNet-TLS algorithm available through **(9)** (https://github.com/DIAGNijmegen/pathology-hooknet-tls)
- Python utility libraries openslide, matplotlib, numpy, as well as libraries needed to train the model using GPUs (both Pytorch and Tensorflow are supported).

---

**Box 1. Content of the MeasureAnnotationAreas.ijm script.**

To create the script, navigate to ImageJ/Plugins/New/Macro, paste this content and save the script file.

```
run ("Clear Results");
dir = getDirectory ("Choose destination Directory ");
name = getTitle;
rename ("original");
run ("Set Measurements...", "area shape feret's display redirect=None decimal=3");
run ("To ROI Manager");
//create a mask
run ("Duplicate...", "title=mask");
rename (name);
run ("8-bit");
run ("Gaussian Blur...", "sigma=2");
setOption ("BlackBackground", false);
//change the AutoThreshold mode if needed (alternatives: "Otsu", "Percentile","Mean", etc)
setAutoThreshold ("Mean");
run ("Create Selection");
run ("Measure");
run ("Convert to Mask"); //mask is 255
run ("Select None");
// loop through Annotations (ROIs)
for (m = 0; m < roiManager ("count"); m++){
selectWindow (name);
roiManager ("Select", m);
run ("Measure");
}
print ("processing: "+name);
saveAs ("results", dir+name+"_results.csv");
selectWindow (name);
saveAs ("tiff", dir+name+"_mask.tiff");
run ("Close All");
```

---

**Box 2. Content of the Summarize.Rmd script.**

To create the script, navigate to RStudio/File/New file/R Markdown, paste this content and save the script file.

```
---
title: "QuPath_annotation_summary"
output: html_document
---

# dependencies
```{r}
library (reshape2)
library (data.table)
```



```
library (dplyr)
```
# load csv data from ImageJ
```{r}
reanalysis_folder <- getwd ()
csv_files <- list.files (path = reanalysis_folder, pattern='results.csv', full=TRUE,
                  ignore.case = TRUE, recursive = TRUE)
 (results = rbindlist (lapply (csv_files, fread)))
```
# add columns
```{r}
#find patterns and replace them to get new labels scan id and annotations
results[,scan_id:=gsub (pattern = "^ (.*ndpi).*", replacement = "\\1", x = Label)]
#results[,scan_id:=gsub (pattern = "^ (.*svs).*", replacement = "\\1", x = Label)]
results[,annotation:=gsub ("^.*:","",Label)]
#for empty annotations
results[annotation==Label, annotation:="total area"]
```
# check results file
```{r}
results
unique (results$annotation)
```
# generate a summary table
```{r}
 (summary_table <- dcast.data.table (results,scan_id ~annotation, value.var = "Area", fun.aggregate = list (sum,mean,length)))
```
# add percentages
```{r}
#select variables of interest for which to compute densities
#adjust names as needed for your study
TLS_vars <- colnames (summary_table)[names (summary_table)%like%c ("TLS")]
GC_vars <- colnames (summary_table)[names (summary_table)%like%c ("GC")]
vars <- c (TLS_vars,GC_vars)
summary_table[, paste0 (names (summary_table)[which (names (summary_table) %in% vars)], '_Density') := lapply (.SD, function (x)
      x/`Area_sum_total area`), .SDcols = which (names (summary_table) %in% vars)]
```
#save file
```{r}
write.csv (summary_table, file = file.path (reanalysis_folder, "QuPath_Annotations_summary.csv"))
```

**Box 3. Content of the json_export.groovy script.**

To create the script, navigate to QuPath/Automate/Script editor, paste this content and save the script file.

```
import java.io.FileWriter;
import qupath.lib.scripting.QP
import qupath.lib.io.GsonTools

def output_folder = 'Volumes/.../'
def image_extension = '.svs'

def imageData = QPEx.getCurrentImageData ()
def output_path = output_folder + imageData.toString ().split ()[-1].split (image_extension)[0]+'.json'

def annotations = getAnnotationObjects ()
boolean prettyPrint = true
def gson = GsonTools.getInstance (prettyPrint)
def json =  gson.toJson (annotations)

FileWriter file = new FileWriter (output_path)
file.write (json)
file.flush ()
file.close ()
print ('done')
```



## 3. Methods

### 3.1. Creating manual annotations in histology H&E images

1. Include images where tumor invasive front and adjacent normal tissue is available to ensure the most likely sample material for TLS identification. Include images that lack the above-mentioned regions only if they have dense lymphocytic aggregates.
2. Create a QuPath project with all selected images of interest.
3. Generate manual annotations using the drawing tools from the top menu of all *de novo* formed lymphoid aggregates, homeostatic MALT (if pertinent), LN (if pertinent) and other regions describing non-lymphoid parenchyma (REST) important for your study in each image (**Figure 3**, step 1). Save changes via QuPath/File/Save.
4. Extract the annotation data from each image (**Figure 3** steps 2-4):
   - Double left click anywhere in the image outside of an annotation to deselect any annotation that might be selected (otherwise only this region will be exported to ImageJ).
   - Click on the ImageJ button on the top right corner of the QuPath window.
   - Select the Send region to ImageJ option.
   - Indicate the image resolution (available under the Image tab near the QuPath Project tab) in the Resolution window and tick the Include ROI and Include overlay options.
   - The image with the marked annotations will open within the ImageJ software automatically.
   - Navigate to ImageJ/Plugins/Macros toolbar and click Run.
   - Choose the MeasureAnnotationAreas.ijm script file and follow the prompts to save the extracted annotation data in a dedicated folder.
   - Alternatively to Run, click Edit to make changes to the MeasureAnnotationAreas.ijm script if needed.
   - Close the ImageJ window and return to the QuPath project.

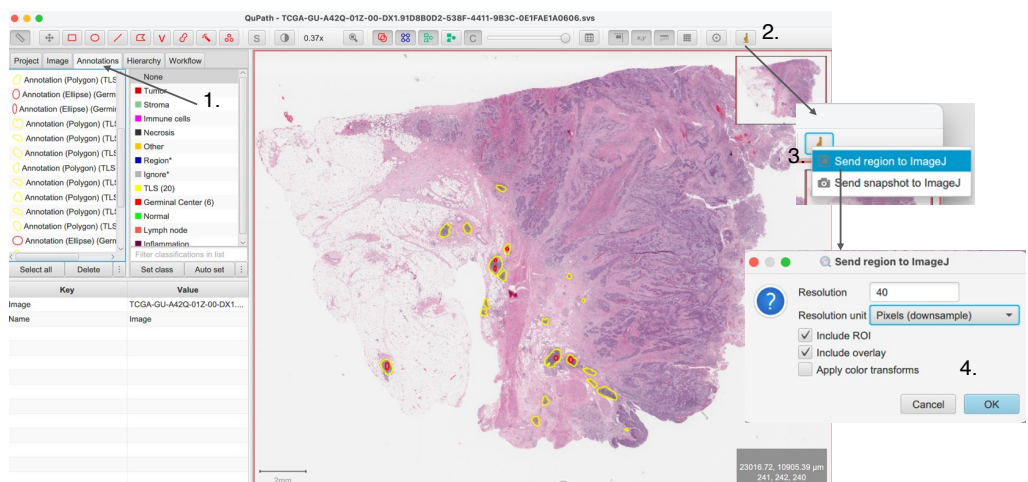

**Figure 3. Generation of manual annotations in H&E images using QuPath.**

1. generate annotations corresponding to tissue regions of interest. 2. Export annotations to ImageJ for automated quantification by clicking on the ImageJ icon in to top toolbar. 3. Indicate the image resolution and inclusion parameters. 4. Confirm choice and continue within ImageJ, which will open automatically.



5. Export the annotation information for training the HookNet-TLS algorithm (**Note 3**):
    o Navigate to QuPath/Automate/Script Editor.
    o Open json_export.groovy file via then QuPath/File/Open.
    o When using the script for the first time, edit the following details in the script: (1) type in the intended output_folder (line 5 of the script), (2) indicate the file format of the images (.svs, .ndpi, etc.) in image_extension (line 6), and (3) save the script.
    o Export the manual annotations in a .json format via QuPath /Run/Run.
    o Close the Script Editor window and return to the QuPath project to annotate the next image.
6. After all images in the project have been processed following steps 1. to 5., use Rstudio to run the Summarize.Rmd script and generate a final data table containing all manual annotation data. Follow steps described in the script and use either the annotation count or the total measured annotation area divided by total tissue area to obtain an aggregate and/or germinal center density value of each image to perform normalized comparisons across tissues/patients/animals/groups.

3.2. Training the HookNet-TLS deep learning algorithm

The HookNet-TLS pipeline is written in Python and is available on GitHub: https://github.com/DIAGNijmegen/pathology-hooknet-tls/tree/main. This code repository contains instructions on how to check out and run a pre-trained HookNet-TLS model using Docker (i.e., a virtual application container based on an operating system and additional software tools), as detailed in the Installation section of this repository. But it also points to the two frameworks that are the foundation of HookNet-TLS: the original HookNet software package *(32)*. It contains the model definition and the WholeSlideData package with features to read, sample from and write to whole-slide images and corresponding manual annotations. In order to train HookNet-TLS with manual annotations and images from Section 3.1, follow the steps described below. An example of all the steps required to train a HookNet model, including its incarnation HookNet-TLS, is contained in the Jupyter Notebook named "HookNetPracticalGuide.ipynb" in the notebooks directory of the HookNet repository.

### 3.2.1 Preparation
1. Install HookNet as described here: https://github.com/DIAGNijmegen/pathology-hooknet
2. Install WholeSlideData as described here (**Note 4**): https://github.com/DIAGNijmegen/pathology-whole-slide-data
3. Install the Python utility libraries openslide, matplotlib, numpy, as well as the library for model training using GPUs Pytorch or Tensorflow.



### 3.2.2 Training

1. **Config file**. While the training routine follows the one of the general HookNet model, HookNet-TLS has some model- and applications-specific differences. Those should be configured, prior to model training, in a config.yaml file (see an example here: https://github.com/DIAGNijmegen/pathology-hooknet/blob/master/notebooks/user_config.yml. In particular, HookNet-TLS takes input data at 0.5 and 2.0 um/px magnification (**Note 5**), uses 32 filters per UNet block (to reduce the number of parameters in the model from 50M in the original HookNet to 25M in HookNet-TLS), and based on annotations described in 3.1, takes three classes annotated as TLS, GC and REST, with REST including non-lymphoid and non-GC regions. Note that while training, regions containing annotations of MALT and LN are ignored. Assuming that those classes are also annotated in the test set, they will be ignored in the inference as well (see Section 3.2.3). Examples of some of these settings can be found in the configuration file used for inference: https://github.com/DIAGNijmegen/pathology-hooknet-tls/blob/main/hooknettls/configs/config.yml.
2. **Data file**. The list of training WSIs and their manual annotations should be detailed in the data.yaml file, which is also input to the training routine. An example of such a file (only containing one training image) can be found here: https://github.com/DIAGNijmegen/pathology-hooknet/blob/master/notebooks/data.yml
3. **Training the model**. Once all these preparatory steps have been complete, run the actual code to train the model. Examples are detailed in sections 4.1 and 4.2 of the example notebook (both pytorch and tensorflow). The trained model will be finally stored in the file hooknet_weights.h5.

### 3.2.3 Inference

Once the model is trained, it can be tested on unseen data to perform inference, i.e., segment and detect TLS and germinal centers.

1. **Running the "official" pre-trained HookNet-TLS**. Should you be interested in performing inference on your data using the model presented in **(9)**, you can run the hooknettls application via docker as explained in the Usage section here: https://github.com/DIAGNijmegen/pathology-hooknet-tls/tree/main
   ```
   python3 -m hooknettls \
       hooknettls.default.image_path="image-path-goes-here" \
       hooknettls.default.mask_path="tissue-mask-path-goes-here"
   ```

   The hooknettls application needs a tissue mask as input, in the form of a binary mask in a multiresolution image file format, to exclude applying the model to the image background. When manual annotations of MALT and LN are available, those regions will be added to the background class, and therefore ignored during inference.
2. **Running your own (custom-trained) model**. To perform inference with the model that you have trained, you can follow the steps described in sections 4.3 and 4.4 of the



example notebook. This will produce the segmentation output, which needs to be post-processed to apply the filtering and the detection criteria explained in **(9)**. Functionalities to apply post-processing are implemented in the file https://github.com/DIAGNijmegen/pathology-hooknet-tls/blob/main/hooknettls/postprocessing.py.

3. **Packing your model in the HookNet-TLS docker**. If you want to use your trained model as part of the hooknettls application, which also includes post-processing, you can build the hooknettls docker using the Dockerfile from the HookNet-TLS repo (https://github.com/DIAGNijmegen/pathology-hooknet-tls/blob/main/Dockerfile) and point to your model weights in the config file: https://github.com/DIAGNijmegen/pathology-hooknet-tls/blob/main/hooknettls/configs/config.yml. The model weights can also be found in https://zenodo.org/records/10614942.

## 4. Notes

1. The 40X scanning resolution allows in-depth exploration of tissue morphology, which can help in deciding whether a lymphoid aggregate contains follicular stromal cells and potential germinal center cells; Nevertheless, 20X scans may also be sufficient for annotations of lymphoid aggregates. It is important to note the scanning resolution for every image when exporting the annotations to ImageJ (see step 3.1.4).
2. H&E or H&E with Saffron can be used. However, the published HookNet-TLS was only trained with H&E images and was not tested on H&E with Saffron.
3. Alternatively, a geojson export function in QuPath can be used, which does not require the script, as described here: https://qupath.readthedocs.io/en/stable/docs/advanced/exporting_annotations.html The output of the geojson function is supported in WholeSlideData.
4. The data reading and writing using WholeSlideData supports multiple backends such as openslide, pyvips and ASAP.
5. The use of two branches at 0.5 and 2.0 um/px is a specific design choice of HookNet-TLS. While training the model, you can opt for more branches and/or different resolutions as these are hyperparameters in the workflow.

37. Zhang T, Lei X, Jia W, et al (2023) Peritumor tertiary lymphoid structures are associated with infiltrating neutrophils and inferior prognosis in hepatocellular carcinoma. Cancer Med 12:3068–3078
14